\documentclass[runningheads]{llncs}
\usepackage[T1]{fontenc}
\usepackage[colorlinks, linkcolor=blue, citecolor=blue,urlcolor=blue]{hyperref}
\usepackage{amsmath}
\usepackage{amssymb}
\usepackage[misc]{ifsym}
\usepackage{graphicx,verbatim}
\usepackage{lipsum}
\usepackage{booktabs}
\usepackage{adjustbox}
\usepackage{multirow}
\usepackage{array}
\usepackage{tabularx}
\usepackage{esvect}
\begin{document}
\title{tCURLoRA: Tensor CUR Decomposition Based Low-Rank Parameter Adaptation and Its Application in Medical Image Segmentation}
\titlerunning{Tensor CUR Decomposition Based Low-Rank Parameter Adaptation}

\author{Guanghua He\inst{1,2}\and
Wangang Cheng\inst{2} \and
Hancan Zhu\inst{2}\textsuperscript{(\Letter)} \and
Xiaohao Cai\inst{3} \and
Gaohang Yu\inst{1}\textsuperscript{(\Letter)}}

\authorrunning{G. He et al.}
%
\institute{Department of Mathematics, Hangzhou Dianzi University, Hangzhou, 310018, China\\ \email{maghyu@163.com} \and 
School of Mathematics, Physics and Information, Shaoxing University, Shaoxing, 312000, China\\ \email{hancanzhu@yeah.net}\and
School of Electronics and Computer Science, University of Southampton, Southampton, SO17 1BJ, UK}

\maketitle              

\begin{abstract}
Transfer learning, by leveraging knowledge from pre-trained models, has significantly improved the performance of downstream tasks. However, as deep neural networks continue to scale, full fine-tuning poses substantial computational and storage challenges in resource-constrained environments, limiting its practical adoption. To address this, parameter-efficient fine-tuning (PEFT) methods have been proposed to reduce computational complexity and memory requirements by updating only a small subset of parameters. Among them, matrix decomposition-based approaches such as LoRA have shown promise, but often struggle to fully capture the high-dimensional structural characteristics of model weights. In contrast, high-order tensors offer a more natural representation of neural network parameters, enabling richer modeling of multi-dimensional interactions and higher-order features. In this paper, we propose tCURLoRA, a novel fine-tuning method based on tensor CUR decomposition. By stacking pre-trained weight matrices into a third-order tensor and applying tensor CUR decomposition, our method updates only the compressed tensor components during fine-tuning, thereby substantially reducing both computational and storage costs. Experimental results show that tCURLoRA consistently outperforms existing PEFT approaches on medical image segmentation tasks. The source code is publicly available at: \href{https://github.com/WangangCheng/t-CURLora}{https://github.com/WangangCheng/t-CURLora}.


\end{abstract}
\renewcommand{\thefootnote}{}
\footnotetext{G. He and W. Cheng---Contributed equally.}
\section{Introduction}
Transfer learning has significantly improved the performance of target tasks, especially in data-scarce scenarios, by leveraging knowledge from pre-trained models \cite{5288526,weiss2016survey}. However, as deep learning models continue to scale, full fine-tuning incurs substantial computational and storage costs in resource-constrained environments, limiting its practical applicability. To address this challenge, parameter-efficient fine-tuning (PEFT) methods have been introduced. By reducing the number of parameters that need to be updated in deep neural networks (DNNs), PEFT methods effectively lower computational complexity and storage demands, and have emerged as a prominent research focus in recent years \cite{ding2023parameter,han2024parameter,hu2021lora}.

Low-Rank Adaptation (LoRA) is a well-established PEFT method that reduces the number of trainable parameters by introducing low-rank matrices to incrementally update pre-trained weights in DNNs, while maintaining high model performance \cite{hu2021lora}. To further improve generalization, Hydra extends this idea through a multi-head low-rank adaptation strategy, combining parallel and sequential branching structures to enhance model expressiveness \cite{KIM2024106414}. PiSSA, built upon singular value decomposition (SVD), improves fine-tuning efficiency by focusing on the dominant singular values and their corresponding vectors \cite{NEURIPS2024_db36f4d6}. In parallel, both CURLoRA and PMSS leverage CUR matrix decomposition to further optimize adaptation. CURLoRA mitigates catastrophic forgetting during continuous learning \cite{fawi2024curlora}, whereas PMSS improves adaptability to complex tasks by selecting skeletal substructures from pre-trained weight matrices \cite{wang2024pmss}.

Current PEFT methods based on low-rank adaptation predominantly rely on matrix decomposition. However, such approaches often struggle to capture the high-dimensional structural properties of DNN weights. In contrast, high-order tensors offer a more natural and expressive representation, enabling the modeling of complex multi-dimensional interactions and higher-order feature dependencies \cite{9420085}. Incorporating tensor decomposition into PEFT has the potential to not only improve the efficiency of transfer learning, but also to reveal latent structures embedded within high-dimensional weight representations.

In this paper, we propose tCURLoRA, a key substructure fine-tuning method based on tensor CUR decomposition. Specifically, we stack pre-trained weight matrices from multiple layers along the frontal dimension to form a third-order tensor, which captures shared architectural patterns across transformer layers. This tensorized structure facilitates the application of tensor CUR decomposition, enabling more effective modeling of cross-layer correlations than traditional matrix-based approaches. During fine-tuning, only the compressed tensor components are updated, substantially reducing the number of trainable parameters and minimizing both computational and memory overhead. The main contributions of this work are as follows:
\begin{itemize}
    \item We propose tCURLoRA, which leverages tensor CUR decomposition on stacked pre-trained weights to exploit high-dimensional structures and enable efficient adaptation by updating only the most informative components.

    \item Experimental results on three transfer learning tasks demonstrate that tCURLoRA consistently outperforms existing PEFT baselines in segmentation accuracy under limited data conditions.
\end{itemize}

\section{Method}
\subsection{Overall Architecture}
\begin{figure}
    \includegraphics[width=\textwidth]{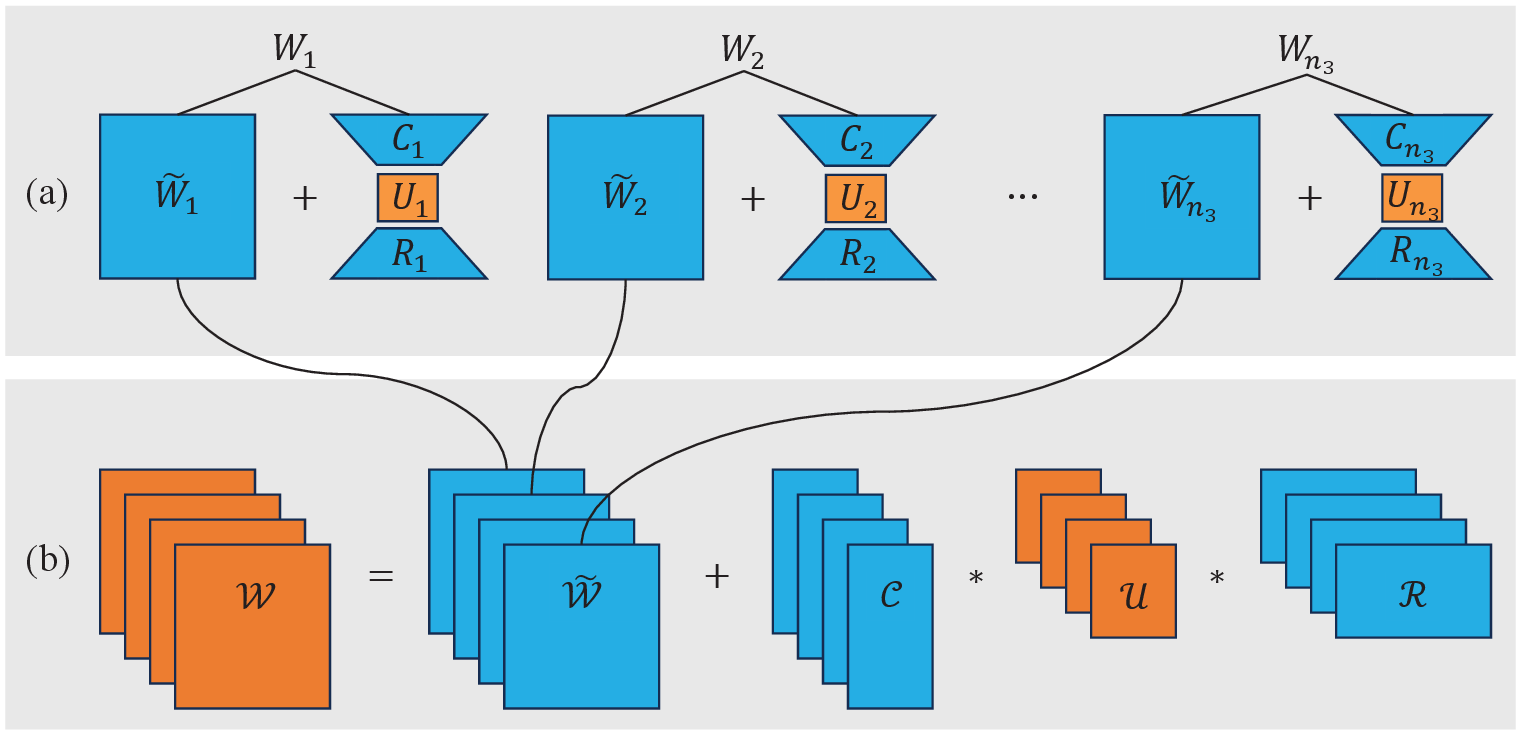}
    \caption{Comparison of PEFT Methods: Matrix vs. Tensor CUR Decomposition. (a) Matrix CURLoRA fine-tunes weight matrices independently. (b) tCURLoRA stacks weight matrices into a 3D tensor for fine-tuning. Blue indicates frozen parameters, orange indicates updated parameters, and * denotes the tensor product.} \label{fig1}
\end{figure}

Figure \ref{fig1}(a) illustrates a PEFT method based on matrix CUR decomposition \cite{fawi2024curlora,wang2024pmss}. In this approach, matrix CUR decomposition is independently applied to each pre-trained weight matrix for fine-tuning. Let $\widetilde{W}_i \in \mathbb{R}^{n_1 \times n_2}$, $i = 1, 2, \dots, n_3$ denote the $n_3$ pre-trained weight matrices. During fine-tuning, these matrices remain frozen, while their increments $\Delta W_i$ are updated. The update rule is given by:

\begin{equation}
W_i = \widetilde{W}_i + \Delta W_i = \widetilde{W}_i + C_i U_i R_i,\quad i = 1, 2, \ldots, n_3,
\end{equation}
where $C_i \in \mathbb{R}^{n_1 \times c}$ contains $c$ sampled columns from $\widetilde{W}_i$, and $R_i \in \mathbb{R}^{r \times n_2}$ contains $r$ sampled rows. Both $C_i$ and $R_i$ remain fixed during fine-tuning, while $U_i \in \mathbb{R}^{c \times r}$ is initialized to zero and optimized. To simplify hyperparameter tuning, $c$ and $r$ are typically set equal \cite{fawi2024curlora,wang2024pmss}.

Figure \ref{fig1}(b) shows the proposed tCURLoRA, which extends matrix CUR to the tensor setting in order to better exploit high-dimensional structural patterns. Specifically, pre-trained weight matrices are stacked along the frontal dimension to form a third-order tensor $\widetilde{\mathcal{W}} \in \mathbb{R}^{n_1 \times n_2 \times n_3}$. Fine-tuning is performed via tensor CUR decomposition, expressed as:

\begin{equation}\label{eq2}
\mathcal{W} = \widetilde{\mathcal{W}} + \Delta \mathcal{W} = \widetilde{\mathcal{W}} + \mathcal{C} * \mathcal{U} * \mathcal{R},
\end{equation}
where $\mathcal{C} \in \mathbb{R}^{n_1 \times r \times n_3}$ and $\mathcal{R} \in \mathbb{R}^{r \times n_2 \times n_3}$ are fixed low-rank tensors obtained from the tensor CUR decomposition of $\widetilde{\mathcal{W}}$, and $\mathcal{U} \in \mathbb{R}^{r \times r \times n_3}$ is a learnable tensor initialized to zero. The symbol “$*$” denotes the tensor product. Here, $r$ represents the number of sampled rows and columns shared across slices.

\subsection{Tensor CUR Decomposition}

The tensor CUR decomposition adopted in this work is based on the tensor product (t-product) framework \cite{kilmer2013third,liu2022tensor}.

\begin{definition}
Let $\mathcal{A} \in \mathbb{R}^{n_1 \times n_2 \times n_3}$ and $\mathcal{B} \in \mathbb{R}^{n_2 \times l \times n_3}$ be two third-order tensors. The t-product $\mathcal{A} * \mathcal{B}$ yields a tensor of size $n_1 \times l \times n_3$, defined as:
\[
\mathcal{A} * \mathcal{B} = \mathrm{fold}\left(\mathrm{circ}(\mathcal{A})\mathrm{MatVec}(\mathcal{B})\right),
\]
where the operators $\mathrm{circ}$, $\mathrm{MatVec}$, and $\mathrm{fold}$ are described below.
\end{definition}

The expressions for $\mathrm{circ}(\mathcal{A})$ and $\mathrm{MatVec}(\mathcal{B})$ are given by:
\[
\operatorname{circ}(\mathcal{A}) =
\begin{bmatrix}
A_1 & A_{n_3} & \cdots & A_2 \\
A_2 & A_1 & \cdots & A_3 \\
\vdots & \vdots & \ddots & \vdots \\
A_{n_3} & A_{n_3-1} & \cdots & A_1
\end{bmatrix}, \quad
\operatorname{MatVec}(\mathcal{B}) =
\begin{bmatrix}
B_1 \\
B_2 \\
\vdots \\
B_{n_3}
\end{bmatrix},
\]
where $A_i = \mathcal{A}(:,:,i)$ and $B_i = \mathcal{B}(:,:,i)$ denote the $i$th frontal slices of tensors $\mathcal{A}$ and $\mathcal{B}$, respectively. The $\mathrm{fold}$ operation transforms the matrix $\mathrm{MatVec}(\mathcal{B})$ back into the original tensor $\mathcal{B}$, i.e., $\mathrm{fold}(\mathrm{MatVec}(\mathcal{B})) = \mathcal{B}$.

It is known that a block circulant matrix can be diagonalized via the Discrete Fourier Transform (DFT) \cite{golub2013matrix}, i.e., $(F \otimes I_{n_1}) \mathrm{circ}(\mathcal{A}) (F^* \otimes I_{n_2})$ results in a block-diagonal matrix, where $F \in \mathbb{R}^{n_3 \times n_3}$ is the DFT matrix, $F^*$ is its conjugate transpose, $\otimes$ denotes the Kronecker product, and $I_{n_1}$ is the $n_1 \times n_1$ identity matrix. In addition, we have:
\[
\mathrm{circ}(\mathcal{A}) \mathrm{MatVec}(\mathcal{B}) = (F^* \otimes I_{n_1}) \left( (F \otimes I_{n_1}) \mathrm{circ}(\mathcal{A}) (F^* \otimes I_{n_2}) \right) (F \otimes I_{n_2}) \mathrm{MatVec}(\mathcal{B}).
\]
 Thus, the t-product can be efficiently computed as follows:  
(i) apply the DFT along the third dimension of tensors $\mathcal{A}$ and $\mathcal{B}$ to transform them into the frequency domain;  
(ii) perform matrix multiplication on corresponding frontal slices;  
(iii) apply the inverse DFT along the third dimension of the resulting tensor to obtain the final output.

The tensors $\mathcal{C}$ and $\mathcal{R}$ in Equation~\eqref{eq2} are derived via the tensor CUR decomposition of $\widetilde{\mathcal{W}}$, following the procedure in \cite{Chen26042022}. Specifically, we first apply the Fast Fourier Transform (FFT) along the third dimension of $\widetilde{\mathcal{W}} \in \mathbb{R}^{n_1 \times n_2 \times n_3}$ to obtain
\[
\widehat{\mathcal{W}} = \mathrm{fft}(\widetilde{\mathcal{W}}, [], 3).
\]
To evaluate the importance of columns, we define a column score $\alpha_j$ as:
\begin{equation}
\alpha_j = \frac{\sum_{k=1}^{n_3} \|\widehat{\mathcal{W}}(:, j, k)\|_2}{\sum_{j=1}^{n_2} \sum_{k=1}^{n_3} \|\widehat{\mathcal{W}}(:, j, k)\|_2}, \quad j = 1, 2, \ldots, n_2,
\end{equation}
where $\|\cdot\|_2$ denotes the $\ell_2$-norm of a vector. The top $r$ columns with the largest scores are selected to form index set $J$. Similarly, we define the row score $\beta_i$ as:
\begin{equation}
\beta_i = \frac{\sum_{k=1}^{n_3} \|\widehat{\mathcal{W}}(i, J, k)\|_2}{\sum_{i=1}^{n_1} \sum_{k=1}^{n_3} \|\widehat{\mathcal{W}}(i, J, k)\|_2}, \quad i = 1, 2, \ldots, n_1.
\end{equation}
The top $r$ rows form index set $I$. Using sets $I$ and $J$, we extract:
\[
\widehat{\mathcal{C}} = \widehat{\mathcal{W}}(:, J, :), \quad
\widehat{\mathcal{U}} = \widehat{\mathcal{W}}(I, J, :), \quad
\widehat{\mathcal{R}} = \widehat{\mathcal{W}}(I, :, :).
\]
Finally, we apply the inverse FFT along the third dimension to obtain the final decomposition components:
\[
\mathcal{C} = \mathrm{ifft}(\widehat{\mathcal{C}}, [], 3), \quad
\mathcal{U} = \mathrm{ifft}(\widehat{\mathcal{U}}, [], 3), \quad
\mathcal{R} = \mathrm{ifft}(\widehat{\mathcal{R}}, [], 3),
\]
leading to the approximation $\widetilde{\mathcal{W}} \approx \mathcal{C} * \mathcal{U}^\dagger * \mathcal{R}$, where $\mathcal{U}^\dagger$ denotes the Moore–Penrose inverse of $\mathcal{U}$.
 
 \subsection{tCURLoRA for Fine-Tuning UNETR Parameters}

We apply tCURLoRA to fine-tune the UNETR network \cite{Hatamizadeh_2022_WACV}, which comprises a Transformer-based encoder with 12 layers and a convolutional decoder, connected by deconvolution and convolution layers. Given that most parameters reside in the Transformer modules, tCURLoRA fine-tunes the Transformer modules while fully updating the decoder parameters and keeping the skip connection parameters frozen.

Each Transformer layer consists of two primary components: Multi-Head Self-Attention (MHSA) and a Multi-Layer Perceptron (MLP). In the MHSA module, the queries, keys, values, and outputs are generated using projection matrices $W_{a}, W_{k}, W_{v}, W_{o} \in \mathbb{R}^{d \times d}$. These are divided into $N_h$ attention heads, where each head uses its own set of matrices $W_{q}^{(i)}, W_{k}^{(i)}, W_{v}^{(i)}, W_{o}^{(i)} \in \mathbb{R}^{d \times d_h}$, with $d_h = d / N_h$ and $i = 1, 2, \ldots, N_h$. The output of the MHSA module is computed as:
\[
MHSA(X) = \sum_{i=1}^{N_h} \mathrm{softmax} \left( \frac{X W_{q}^{(i)} W_{k}^{(i)^T} X^T}{\sqrt{d_h}} \right) X W_{v}^{(i)} W_{o}^{(i)^T}.
\]
The MLP module contains two fully connected layers (biases omitted), expressed as:
\[
MLP(X) = \mathrm{GELU}(X W_{up}) W_{down},
\]
where $\mathrm{GELU}$ is the Gaussian Error Linear Unit activation function \cite{hendrycks2016gaussian}, and $W_{up} \in \mathbb{R}^{d \times 4d}$, $W_{down} \in \mathbb{R}^{4d \times d}$ are the weights of the two layers.

Each Transformer layer therefore contains four $d \times d$ matrices from the MHSA module and two matrices—$d \times 4d$ and $4d \times d$—from the MLP module. Across 12 layers, this results in 48 $d \times d$ matrices, 12 $d \times 4d$ matrices, and 12 $4d \times d$ matrices. These are concatenated along the frontal dimension to form three tensors: $\mathcal{W}_{sa} \in \mathbb{R}^{d \times d \times 48}$, $\mathcal{W}_{up} \in \mathbb{R}^{d \times 4d \times 12}$, and $\mathcal{W}_{down} \in \mathbb{R}^{4d \times d \times 12}$. In tCURLoRA, these tensors are fine-tuned independently to enable efficient model optimization.

\section{Experiments}
\subsection{Datasets}
The UNETR model is first pre-trained on the BraTS2021 dataset \cite{baid2021rsna}, which contains multimodal magnetic resonance imaging (MRI) data, including T1, T1ce, T2, and FLAIR modalities. For each patient, the image resolution is $240 \times 240 \times 155$ with a voxel size of $1 \times 1 \times 1$ mm$^3$. The dataset provides detailed tumor annotations for 1,251 cases, segmented into three regions: enhancing tumor (ET), peritumoral edema/infiltrative tissue (ED), and necrotic tumor core (NCR).

We then apply the tCURLoRA method to transfer the pre-trained UNETR segmentation model to three downstream datasets: EADC-ADNI \cite{boccardi2015training}, LPBA40 \cite{SHATTUCK20081064}, and UPENN-GBM \cite{bakas2021multi}.
The EADC-ADNI dataset, derived from the ADNI database, includes MRI scans of 135 patients with a resolution of $197 \times 233 \times 189$ and a voxel size of $1 \times 1 \times 1$ mm$^3$. The ADNI project, launched in 2003, was initially designed to assess biomarkers for tracking mild cognitive impairment (MCI) and early Alzheimer's disease (AD), and now focuses on validating biomarkers for clinical trials and enhancing data diversity (\url{http://adni.loni.usc.edu/}).
Hippocampal annotations are based on the harmonized segmentation protocol proposed by the European Alzheimer’s Disease Consortium and ADNI \cite{boccardi2015training} (\url{http://www.hippocampal-protocol.net}). During quality control, five cases with annotation inconsistencies were identified and excluded to ensure labeling accuracy.

The LPBA40 dataset, developed by the Laboratory of Neuroimaging (LONI), comprises 3D brain MRI scans from 40 healthy adults, with a resolution of $256 \times 124 \times 256$ and voxel dimensions of $0.8938 \times 1.500 \times 0.8594$ mm$^3$. It provides detailed manual annotations for 56 brain tissues and structures. In this study, only hippocampal annotations are used to evaluate the proposed method. 

The UPENN-GBM dataset contains MRI scans from 630 glioblastoma patients collected at the University of Pennsylvania Health System and made publicly available via the Cancer Imaging Archive (TCIA) \cite{bakas2021multi}. All scans were acquired pre-operatively using a 3T MRI scanner, with imaging modalities including T1, T2, contrast-enhanced T1 (T1GD), and FLAIR, along with corresponding segmentation labels. A subset of 147 scans was manually annotated by clinical experts to delineate three tumor sub-regions: necrotic core (NC), peritumoral edema (ED), and enhancing tumor (ET). In this study, we use these 147 expert-annotated cases, selecting the T1GD modality and defining the Whole Tumor (WT) region—comprising all three sub-regions—as the segmentation target.

All images from the aforementioned datasets were skull-stripped and registered to the MNI152 standard space, resulting in a uniform voxel size of $1 \times 1 \times 1$ mm$^3$.

\subsection{Experimental Details}

We conducted experiments using PyTorch on two NVIDIA GeForce RTX 4090D GPUs. During pre-training, we merged the three annotated tumor sub-regions in the BraTS2021 dataset into a single tumor region for binary segmentation, using only the T1ce modality. We split the 1,251 publicly annotated cases into training (1,000), validation (125), and test (126) sets following an 8:1:1 ratio. We configured the hyperparameters according to the UNETR paper \cite{Hatamizadeh_2022_WACV}.

In the fine-tuning phase, we randomly selected five samples from each of the EADC-ADNI and LPBA40 datasets for training, with the remaining samples used for testing. For the UPENN-GBM dataset, 10\% of the samples were used for training and the remaining 90\% for testing. This setting reflects the relatively consistent anatomical structure of the hippocampus compared to the higher variability of brain tumors, thereby simulating different levels of data scarcity.
We trained the models using the Adam optimizer with a batch size of 4 and a polynomial learning rate decay schedule, starting with an initial learning rate of 0.001 and applying a decay factor of 0.9 per iteration. Random cropping of size $128 \times 128 \times 128$ was applied during training. Data augmentation techniques included: (1) random mirroring with a probability of 50\%, (2) intensity shifts within the range $\left[ -0.1, 0.1 \right]$, and (3) scaling within the range $\left[ 0.9, 1.1 \right]$. The network was trained for 1,000 epochs using Dice loss with L2 regularization (weight decay of $10^{-5}$).

During testing, $128 \times 128 \times 128$ patches were extracted using a non-overlapping sliding window strategy. The final segmentation was obtained by averaging the outputs from the last four training epochs. Post-processing was performed to eliminate false positives by identifying connected components and removing those with a volume smaller than 1 cm$^3$, which were considered background.

\subsection{Experimental Results}

We compared the proposed tCURLoRA with several PEFT methods, including Full fine-tuning (Full), Linear probing (Linear), LoRA \cite{hu2021lora}, Adapter \cite{luo2023towards}, SSF \cite{NEURIPS2022_00bb4e41}, LoTR \cite{bershatsky2024lotr}, PISSA \cite{NEURIPS2024_db36f4d6}, and CURLoRA \cite{fawi2024curlora}. For rank-based methods (LoRA, Adapter, LoTR, PISSA, CURLoRA, and tCURLoRA), we tuned the rank $r$ on the EADC-ADNI dataset. The optimal values were $r=32$ for LoRA and Adapter, $r=2$ for LoTR, PISSA, and CURLoRA, and $r=8$ for tCURLoRA. These settings were used in subsequent experiments unless otherwise specified.

\begin{table}[h]
    \centering
    \caption{Comparison of segmentation results of different fine-tuning methods on three datasets, in terms of Dice (\%) and HD95 (mm) metrics. The best results are highlighted in bold.}
    \label{tab1}
    \begin{tabular}{ccccccccc}
        \toprule
        \textbf{Method} & \textbf{\#Params (M)} & \multicolumn{2}{c}{\textbf{EADC-ADNI}} & \multicolumn{2}{c}{\textbf{LPBA40}} & \multicolumn{2}{c}{\textbf{UPENN-GBM}} \\
        \cmidrule(lr){3-4} \cmidrule(lr){5-6} \cmidrule(lr){7-8}
        & & \textbf{Dice} & \textbf{HD95} & \textbf{Dice} & \textbf{HD95} & \textbf{Dice} & \textbf{HD95} \\
        \midrule
        Full & 90.011 & 83.79 & 5.839 & 79.91 & 7.175 & 69.95 & 32.280 \\
        Linear & 59.348 & 83.46 & 5.344 & 80.62 & 6.483 & 72.42 & 29.401 \\
        LoRA~\cite{hu2021lora} & 7.397 & 84.35 & 5.334 & 80.17 & 6.618 & 72.51 & 29.799 \\
        Adapter~\cite{luo2023towards} & 7.987 & 84.08 & 5.663 & 79.72 & 6.793 & 73.46 & 33.357 \\
        SSF~\cite{NEURIPS2022_00bb4e41} & 2.883 & 83.73 & 5.326 & 79.08 & 6.904 & 72.29 & \textbf{28.762} \\
        LoTR~\cite{bershatsky2024lotr} & 2.703 & 84.05 & 5.394 & 80.21 & 7.011 & 71.14 & 32.093 \\
        PiSSA~\cite{NEURIPS2024_db36f4d6} & 2.974 & 84.45 & 5.603 & 80.62 & 6.584 & 72.56 & 31.331 \\
        CURLoRA~\cite{fawi2024curlora} & 2.679 & 84.64 & 5.549 & 79.96 & 6.659 & 72.73 & 29.387 \\
        \textbf{tCURLoRA (ours)} & 2.683 & \textbf{84.95} & \textbf{4.855} & \textbf{81.12} & \textbf{6.305} & \textbf{74.28} & 30.550 \\
        \bottomrule
    \end{tabular}
\end{table}

Table~\ref{tab1} presents the segmentation performance on the EADC-ADNI, LPBA40, and UPENN-GBM datasets. The proposed tCURLoRA method comprises 2.683 million (M) parameters, comparable to SSF, LoTR, PISSA, and CURLoRA, while requiring only 2.98\% of the parameters used in full fine-tuning, thereby significantly improving training efficiency. Compared with full fine-tuning, tCURLoRA improves the Dice coefficient by 1.16\%, 1.21\%, and 4.33\%, and reduces the HD95 metric by 0.984 mm (16.85\%), 0.870 mm (12.13\%), and 1.73 mm (5.36\%) on the three datasets, respectively, indicating enhanced segmentation accuracy. Furthermore, tCURLoRA achieves the highest Dice scores across all datasets, further demonstrating its superior performance in medical image segmentation.

\begin{figure}
    \includegraphics[width=\textwidth]{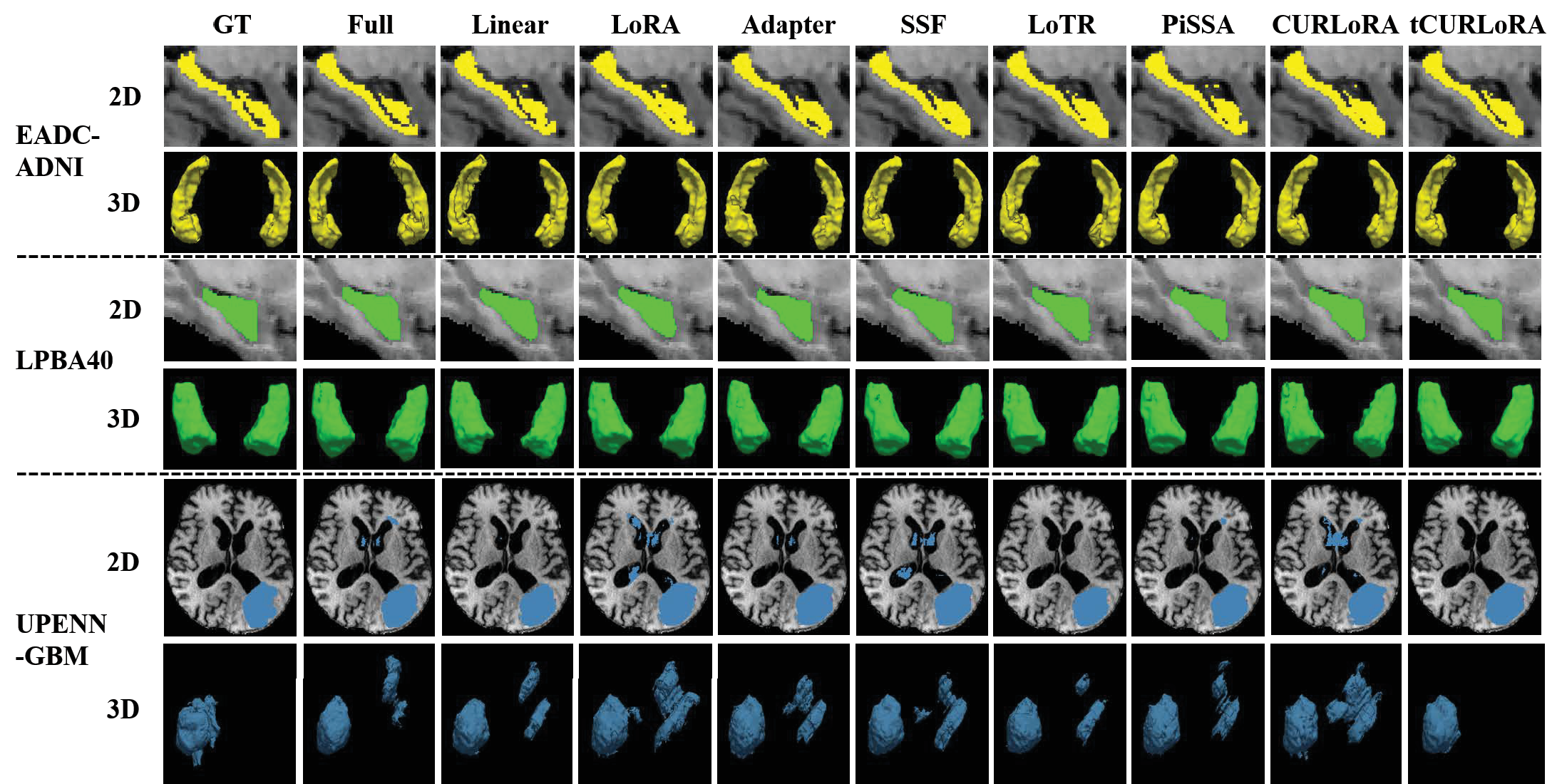}
    \caption{Qualitative Segmentation Results: 2D Slices and 3D Surfaces Across Datasets.} \label{fig2}
\end{figure}

We also evaluated training efficiency in terms of per-epoch runtime and memory consumption. Under identical experimental settings, tCURLoRA achieves 495 ms per epoch and utilizes 11.72 GB of memory, ranking second among all PEFT methods, just behind CURLoRA. In comparison, LoRA requires 562 ms and 15.90 GB, while full fine-tuning takes 684 ms and 18.28 GB. These results underscore the practical advantages of tCURLoRA in resource-constrained environments.
Figure~\ref{fig2} presents qualitative segmentation results, including 2D slices and 3D surface renderings. The predictions generated by tCURLoRA closely match the ground truth (GT), particularly in regions with complex anatomical structures, effectively preserving fine details while minimizing segmentation errors. 

\section{Conclusion}
This study proposes tCURLoRA, a tensor CUR decomposition-based fine-tuning method that improves the efficiency of DNNs, with a focus on medical image segmentation. By reducing learnable parameters and controlling computational complexity, tCURLoRA addresses challenges in training cost, memory usage, and scalability, while maintaining high segmentation accuracy. Future work may explore its extension to broader tasks and datasets, particularly in multimodal and large-scale segmentation scenarios. Further optimization could enhance its adaptability to real-world applications, making tCURLoRA a practical tool for efficient deep learning.
%
%

\begin{credits}
\subsubsection{\ackname} This work was supported by the National Natural Science Foundation of China (No. 12071104) and the Humanities and Social Science Fund of the Ministry of Education of China (23YJAZH232).

\subsubsection{\discintname}
The authors declare that they have no competing financial interests.
\end{credits}
%
%
%
\bibliographystyle{splncs04}
\bibliography{Paper-0014}
%
%
%
%
%
\end{document}